\RequirePackage[hyphens]{url}
\documentclass{article} %
\usepackage{url}

\usepackage[accepted]{icml_conference}
\usepackage{graphicx}
\usepackage{hyperref}
\usepackage{url}
\usepackage{enumitem} %
\usepackage{booktabs, multirow} %
\usepackage{colortbl} %
\usepackage{xspace} %
\usepackage{rotating} %

\usepackage{amsmath,amsfonts,bm}

\def\eqref#1{equation~\ref{#1}}

\def\1{\bm{1}}

\DeclareMathAlphabet{\mathsfit}{\encodingdefault}{\sfdefault}{m}{sl}
\SetMathAlphabet{\mathsfit}{bold}{\encodingdefault}{\sfdefault}{bx}{n}

\newcommand{\R}{\mathbb{R}}

\usepackage{ifthen}  %
\usepackage{color}

\newboolean{set_me_to_remove_todos}

\setboolean{set_me_to_remove_todos}{false} %

\ifthenelse{\boolean{set_me_to_remove_todos}}{
    \newcommand{\pierre}[1]{}
    \newcommand{\robin}[1]{}
    \newcommand{\hady}[1]{}
    \newcommand{\todo}[1]{}
}{
    \newcommand{\pierre}[1]{{\color{blue} [\textbf{Pierre}: #1]}}
    \newcommand{\robin}[1]{{\color{pink} [\textbf{Robin}: #1]}}
    \newcommand{\hady}[1]{{\color{orange} [\textbf{Hady}: #1]}}
    \newcommand{\todo}[1]{{\color{red} [\textbf{TODO}: #1]}}
}

\newcommand{\wm}{\delta} %
\newcommand{\ploss}{TF-Loudness\xspace} 
\newcommand{\ours}{\textrm{AudioSeal}\xspace}

\newcommand{\wavmark}{WavMark\xspace}
\newcommand{\ie}{\textit{i.e.}\xspace}
\newcommand{\eg}{\textit{e.g.}\xspace}

\usepackage{pifont} %

\newcommand{\cmark}{\ding{51}\xspace}%
\newcommand{\xmark}{\ding{55}\xspace}%

\RequirePackage[most]{tcolorbox}
\RequirePackage{xcolor}
\definecolor{metablue}{HTML}{0064E0}
\definecolor{metafg}{HTML}{1C2B33}
\definecolor{metabg}{HTML}{F1F4F7}
\RequirePackage{hyperref}
\RequirePackage[noabbrev,nameinlink]{cleveref}
\hypersetup{
  colorlinks,
  linkcolor=metablue,
  citecolor=metablue,
  urlcolor=metablue
}

\newcommand{\ourtitle}{Proactive Detection of Voice Cloning with Localized Watermarking}
\icmltitlerunning{\ourtitle}

\begin{document}

\twocolumn[
\icmltitle{ 
    \ourtitle
}

\icmlsetsymbol{equal}{*}

\begin{icmlauthorlist}
\icmlauthor{Robin San Roman}{equal,aff1,aff2}
\icmlauthor{Pierre Fernandez}{equal,aff1,aff2} 
\icmlauthor{Hady Elsahar}{equal,aff1} \\
\icmlauthor{Alexandre D\'efossez}{aff3}
\icmlauthor{Teddy Furon}{aff2}
\icmlauthor{Tuan Tran}{aff1} 
\end{icmlauthorlist}

\icmlaffiliation{aff1}{FAIR, Meta}
\icmlaffiliation{aff2}{Inria}
\icmlaffiliation{aff3}{Kyutai}

\icmlcorrespondingauthor{}{robinsr, hadyelsahar, pfz@meta.com}

\icmlkeywords{Speech, Generation, Detection, Watermarking, Voice Cloning}

\vskip 0.3in
]

\printAffiliationsAndNotice{\icmlEqualContribution} %

\begin{abstract}

In the rapidly evolving field of speech generative models, there is a pressing need to ensure audio authenticity against the risks of voice cloning. 
We present \ours, the first audio watermarking technique designed specifically for localized detection of AI-generated speech. 
\ours employs a generator / detector architecture trained jointly with a localization loss to enable localized watermark detection up to the sample level, and a novel perceptual loss inspired by auditory masking, that enables \ours to achieve better imperceptibility. 
\ours achieves state-of-the-art performance in terms of robustness to real life audio manipulations and imperceptibility based on automatic and human evaluation metrics. 
Additionally, \ours is designed with a fast, single-pass detector, that significantly surpasses existing models in speed, achieving detection up to two orders of magnitude faster, making it ideal for large-scale and real-time applications.
Code is available at \href{https://github.com/facebookresearch/audioseal}{github.com/facebookresearch/audioseal}.

\end{abstract}

\section{Introduction}

Generative speech models are now capable of synthesizing voices that are indistinguishable from real ones~\citep{arik2018neural, kim2021conditional, casanova2022yourtts, wang2023neural}.
Though speech generation and voice cloning are not novel concepts, their recent advancements in quality and accessibility have raised new security concerns. 
A notable incident occurred where a deepfake audio misleadingly urged US voters to abstain, showcasing the potential for misusing these technologies to spread false information~\citep{murphy2024biden}.
Regulators and governments are implementing measures for AI content transparency and traceability, including forensics and watermarking -- see \citet{ChineseAIGovernance, EuropeanAIAct, USAIAnnouncement}.

\begin{figure}[t]
    \centering
    \includegraphics[width=0.99\linewidth, clip, trim={0in 1.7in 3.5in 0}]{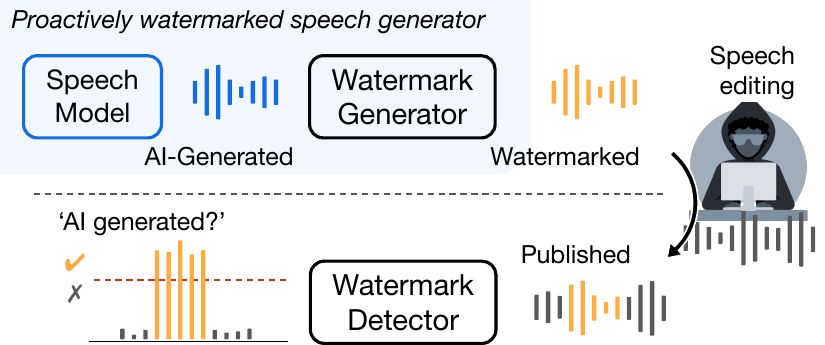}
    \vspace{-0.1cm}
    \caption{
        \textbf{Proactive detection of AI-generated speech.}
        We embed an imperceptible watermark in the audio, which can be used to detect if a speech is AI-generated and identify the model that generated it.
        It can also precisely pinpoint AI-generated segments in a longer audio with a sample level resolution (1/16k seconds).
        }
    \label{fig:fig1}
    \vspace{-0.1in}
\end{figure}

The main forensics approach to detect synthesized audio is to train binary classifiers to discriminate between natural and synthesized audios, a technique highlighted in studies by~\citet{Borsos2022AudioLMAL, Kharitonov2023SpeakRA, le2023voicebox}.
We refer to this technique as \textit{passive detection} since it does not alter of the audio source. 
Albeit being a straightforward mitigation, 
it is prone to fail as generative models advance and the difference between synthesized and authentic content diminishes. 

Watermarking emerges as a strong alternative.
It embeds a signal in the generated audio, imperceptible to the ear but robustly detectable by specific algorithms.
There are two watermarking types: multi-bit and zero-bit. 
Zero-bit watermarking detects the presence or absence of a watermarking signal, which is valuable for AI content detection. 
Multi-bit watermarking embeds a binary message in the content, allowing to link content to a specific user or generative model.
Most deep-learning based audio watermarking methods~\citep{pavlovic2022robust, DEAR_Liu0FMZY23, chen2023wavmark} are multi-bit.
They train a generator to output the watermarked audio from a sample and a message, and an extractor retrieving the hidden message.

Current watermarking methods have limitations.
First, \emph{they are not adapted for detection}.
The initial applications assumed any sound sample under scrutiny was watermarked (\eg IP protection).
As a result, the decoders were never trained on non-watermarked samples.
This discrepancy between the training of the models and their practical use leads to poor or overestimated detection rates, depending on the embedded message (see App.~\ref{app:fpr}).
Our method aligns more closely with the concurrent work by \citet{juvela2023collaborative}, which trains a detector, rather than a decoder.

Second, they \emph{are not localized} and consider the entire audio, making it difficult to identify small segments of AI-generated speech within longer audio clips. 
The concurrent \wavmark's approach~\cite{chen2023wavmark} addresses this by repeating at 1-second intervals a synchronization pattern followed by the actual binary payload. 
This has several drawbacks. 
It cannot be used on spans less than 1 second and is susceptible to temporal edits. 
The synchronization bits also reduce the capacity for the encoded message, accounting for 31\% of the total capacity. 
Most importantly, the brute force detection algorithm for decoding the synchronization bits is prohibitively slow especially on non-watermarked content, as we show in Sec.~\ref{sec:speed}.
This makes it unsuitable for real-time and large-scale traceability of AI-generated content on social media platforms, where most content is not watermarked.

To address these limitations, we introduce \textit{\ours}, a method for localized speech watermarking. 
It jointly trains two networks: a \emph{generator} that predicts an additive watermark waveform from an audio input, and a \emph{detector} that outputs the probability of the presence of a watermark at each sample of the input audio. 
The detector is trained to precisely and robustly detect synthesized speech embedded in longer audio clips by masking the watermark in random sections of the signal.
The training objective is to maximize the detector's accuracy while minimizing the perceptual difference between the original and watermarked audio. 
We also extend \ours to multi-bit watermarking, so that an audio can be attributed to a specific model or version without affecting the detection signal.

We evaluate the performance of \ours to detect and localize AI-generated speech.
\ours achieves state-of-the-art results on robustness of the detection, far surpassing passive detection with near perfect detection rates over a wide range of audio edits.
It also performs sample-level detection (at resolution of 1/16k second), outperforming WavMark in both speed and performance.
In terms of efficiency, our detector is run once and yields detection logits at every time-step, allowing for real-time detection of watermarks in audio streams. 
This represents a major improvement compared to earlier watermarking methods, which require synchronizing the watermark within the detector, thereby substantially increasing computation time.
Finally, in conjunction with binary messages, \ours almost perfectly attributes an audio to one model among $1,000$, even in the presence of audio edits.

Our overall contributions are:
\begin{itemize}[itemsep=1pt, topsep=1pt, leftmargin=*]
\item We introduce \ours, the first audio watermarking technique designed for localized detection of AI-generated speech up to the sample-level;
\item A novel perceptual loss inspired by auditory masking, that enables \ours to achieve better imperceptibility of the watermark signal;  
\item \ours achieves the state-of-the-art robustness to a wide range of real life audio manipulations (section \ref{sec:exps});
\item \ours significantly outperforms the state-of-the-art models in computation speed, achieving up to two orders of magnitude faster detection (section \ref{sec:speed});
\item Insights on the security and integrity of audio watermarking techniques when open-sourcing (section \ref{sec:attacks}).
\end{itemize}

\section{Related Work}

In this section we give an overview of the detection and watermarking methods for audio data. A complementary descrition of prior works 
can be found in the Appendix~\ref{app:related}.

\vspace*{-4pt}\paragraph{Synthetic speech detection.}

Detection of synthetic speech is traditionally done in the forensics community by building features and exploiting statistical differences between fake and real.
These features can be hand-crafted~\cite{sahidullah2015comparison, janicki2015spoofing, albadawy2019detecting, borrelli2021synthetic} and/or learned~\cite{muller2022does, barrington2023single}.
The approach of most audio generation papers~\citep{Borsos2022AudioLMAL, Kharitonov2023SpeakRA, borsos2023soundstorm, le2023voicebox} is to train end-to-end deep-learning classifiers on what their models generate, similarly as \citet{zhang2017investigation}.
Accuracy when comparing synthetic to real is usually good, although not performing well on out of distribution audios (compressed, noised, slowed, etc.).

\vspace*{-4pt}\paragraph{Imperceptible watermarking.} 
Unlike forensics, watermarking actively marks the content to identify it once in the wild. It is enjoying renewed interest in the context of generative models, as it provides a means to track AI-generated content, be it for text~\citep{kirchenbauer2023watermark, aaronson2023watermarking, fernandez2023three},
images~\citep{yu2021responsible, fernandez2023stable, wen2023tree}, or audio/speech~\citep{chen2023wavmark, juvela2023collaborative}.

Traditional methods for audio watermarking relied on embedding watermarks either in the time or frequency domains~\cite{trad_wm_LieC06,trad_wm_KalantariAAA09,trad_wm_NatgunanathanXRZG12,trad_wm_freq_XiangNPHL18,trad_wm_freq_SuZYCJ018,trad_wm_freq_LiuHH19}, usually including domain specific features to design the watermark and its corresponding decoding function.  
Deep-learning audio watermarking methods focus on multi-bit watermarking and follow a generator/decoder framework \cite{tai2019audio, qu2023audioqr, pavlovic2022robust, DEAR_Liu0FMZY23, ren2023speaking}.
Few works have explored zero-bit watermarking~\citep{wu2023adversarial, juvela2023collaborative}, which is better adapted for detection of AI-generated content.
Our rationale is that robustness increases as the message payload is reduced to the bare minimum~\citep{furon2007constructive}.

In this study, we compare our work with the state-of-the-art watermarking method, \wavmark~\cite{chen2023wavmark}, which outperforms previous ones. 
It uses invertible networks to hide 32 bits in 1-second audio segments.
Detection is done by sliding along the audio in 0.05s steps and decoding the message for each window.
If the 10 first decoded bits match a synchronization pattern the rest of the payload is saved (22 bits), and the window can directly slide 1s (instead of the 0.05).
This brute force detection algorithm is prohibitively slow especially when the watermark is absent, since the algorithm will have to attempt and fail to decode a watermark for each sliding window in the input audio (due to the absence of watermark).

\section{Method}
\label{sec:method}
The method jointly trains two models.
The generator creates a watermark signal that is added to the input audio. 
The detector outputs local detection logits.
The training optimizes two concurrent classes of objectives: minimizing the perceptual distortion between original and watermarked audios and maximizing the watermark detection. 
To improve robustness to modifications of the signal and localization, we include a collection of train time augmentations.
At inference time, the logits precisely localize watermarked segments allowing for detection of AI-generated content.
Optionally, short binary identifiers may be added on top of the detection to attribute a watermarked audio to a version of the model while keeping a single detector.

\begin{figure}[b]
    \centering
    \includegraphics[width=\linewidth, clip, trim={0.2 1.8in 2.2in 0}]{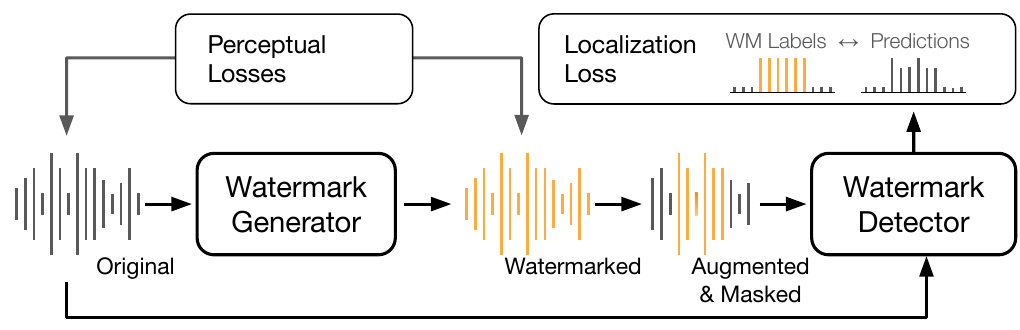}
    \vspace{-0.7cm}
    \caption{Generator-detector training pipeline.}
    \label{fig:method}
\end{figure}

\begin{figure}[b]
    \centering
    \includegraphics[width=0.95\linewidth, clip, trim={0 0 0 0}]{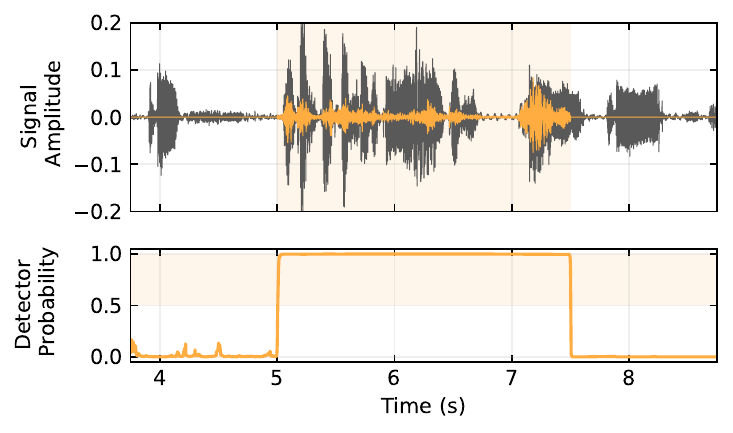}
    \vspace{-0.2cm}
    \caption{
    (Top) A speech signal ({\color{darkgray} gray}) where the watermark is present between 5 and 7.5 seconds ({\color{orange} orange}, magnified by 5).
    (Bottom) The output of the detector for every time step. 
    An {\color{orange!50} orange} background color indicates the presence of the watermark.
    }
    \label{fig:loc_quali}
\end{figure}

\begin{figure*}[t]
    \centering
    \includegraphics[width=0.85\linewidth, clip, trim={0 1.6in 0.8in 0}]{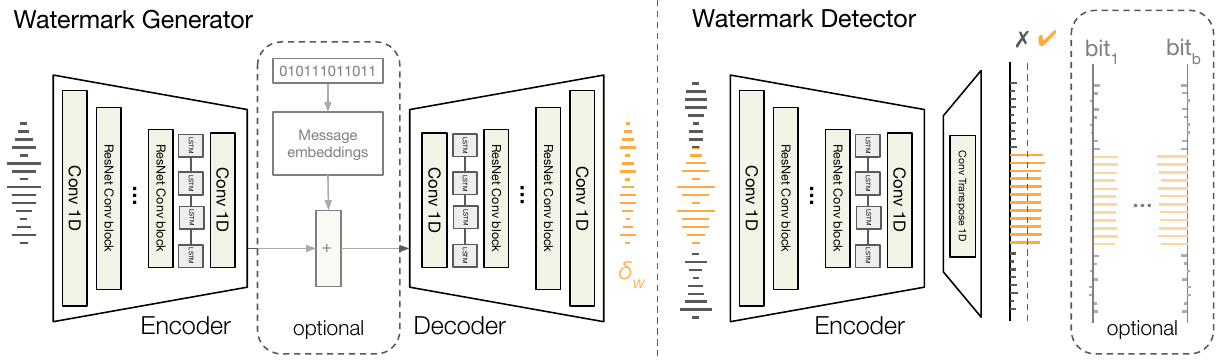}
    \vspace{-0.2cm}
    \caption{
        \textbf{Architectures}. 
        The \emph{generator} is made of an encoder and a decoder both derived from EnCodec's design, with optional message embeddings. 
        The encoder includes convolutional blocks and an LSTM, while the decoder mirrors this structure with transposed convolutions.
        The \emph{detector} is made of an encoder and a transpose convolution, followed by a linear layer that calculates sample-wise logits. 
        Optionally, multiple linear layers can be used for calculating k-bit messages. More details in App. \ref{app:arch}.
    }
    \vspace{-0.2cm}
    \label{fig:archs}
\end{figure*}

\subsection{Training pipeline} 
\autoref{fig:method} illustrates the joint training of the generator and the detector with four critical stages:
\begin{enumerate}[label=(\roman*), itemsep=2pt, topsep=2pt, leftmargin=20pt]
    \item The watermark generator takes as input a waveform $s \in \mathbb{R}^T$ and outputs a watermark waveform $\wm \in \mathbb{R}^T$ of the same dimensionality, where $T$ is the number of samples in the  signal. The watermarked audio is then $s_w = s+\wm$.
    \item To enable sample-level localization, we adopt an augmentation strategy focused on watermark masking with silences and other original audios. This is achieved by 
    randomly selecting $k$ starting points and altering the next $T/2k$ samples from $s_w$ in one of 4 ways: revert to the original audio (\ie $s_w(t)=s(t)$) with probability 0.4; replacing with zeros (\ie $s_w(t)=0$) with probability 0.2; or substituting with a different audio signal from the same batch (\ie $s_w(t) = s'(t)$) with probability 0.2, or not modifying the sample at all with probability 0.2.
    \item The second class of augmentation ensures the robustness against audio editing. 
    One of the following signal alterations is applied: 
    bandpass filter, boost audio, duck audio, echo, highpass filter, lowpass filter, pink noise, gaussian noise, slower, smooth, resample (full details in App.~\ref{app:augmentations}). 
    The parameters of those augmentations are fixed to aggressive values to enforce maximal robustness and the
    probability of sampling a given augmentation is proportional to the inverse of its evaluation detection accuracy.
    We implemented these augmentations in a differentiable way when possible, and otherwise (\eg MP3 compression) with the straight-through estimator~\cite{yin2019understanding} that allows the gradients to back-propagate to the generator.
    \item 
    Detector $D$ processes the original and the watermarked signals, outputting for each a soft decision at every time step, meaning $D(s) \in [0, 1]^T$.
    \autoref{fig:loc_quali}
    illustrates that the detector's outputs
    are at one only when the watermark is present.
\end{enumerate}

The architectures of the models are based on EnCodec~\cite{defossez2022high}.
They are presented in \autoref{fig:archs} and detailed in the appendix~\ref{app:arch}.

\subsection{Losses} 

Our setup includes multiple perceptual losses and a localization loss. 
We balance them during training by scaling their gradients as done by~\citet{defossez2022high}.
The complete list of used losses is detailed bellow.

\vspace{-0.2cm}
\paragraph{Perceptual losses}
enforce the watermark imperceptibility to the human ear.
These include an $\ell_1$ loss on the watermark signal to decrease its intensity, the multi-scale Mel spectrogram loss of \citep{gritsenko2020spectral}, and discriminative losses based on adversarial networks that operate on multi-scale short-term-Fourier-transform
spectrograms. \citet{defossez2022high} use 
this combination of losses for training the EnCodec model for audio compression.

In addition, we introduce a novel time-frequency loudness loss \textbf{\ploss}, which operates entirely in the waveform domain. 
This approach is based on ``auditory masking'', a psycho-acoustic property of the human auditory system already exploited  in the early days of watermarking~\cite{blockrep1-Kirovski2003}:
the human auditory system fails perceiving sounds occurring at the same time and at the same frequency range~\cite{book:audio}.
\ploss is calculated as follows: first, the input signal $s$ is divided into $B$ signals based on non-overlapping frequency bands $s_0, \dots, s_{B-1}$. 
Subsequently, every signal is segmented using a window of size $W$, with an overlap amount denoted by $r$.
This procedure is applied to both the original audio signal \(s\) and the embedded watermark \(\wm\). As a result, we obtain segments of the signal and watermark in time-frequency dimensions, denoted as \(s_b^{w}\) and \(\wm_b^{w}\) respectively.
For every time-frequency window we compute the loudness difference, where loudness is estimated using ITU-R BS.1770-4 recommendations~\cite{loudness} (see App.~\ref{app:loudness} for details):
\vspace{-0.1cm}
\begin{equation}\label{eq:loudness_diff}
    l_b^w = \mathrm{Loudness}(\wm_b^w) - \mathrm{Loudness}(s_b^w).
    \vspace{-0.1cm}
\end{equation}
This measure quantifies the discrepancy in loudness between the watermark and the original signal within a specific time window $w$, and a particular frequency band $b$.
The final loss is a weighted sum of the loudness differences using softmax function:
\vspace{-0.1cm}
\begin{equation}\label{eq:perceptual_loss}
    \mathcal{L}_{loud} = \sum_{b, w} \left(\mathrm{softmax}(l)_b^w * l_b^w\right).
    \vspace{-0.2cm}
\end{equation}
The softmax prevents the model from targeting excessively low loudness where the watermark is already inaudible. 

\vspace{-0.2cm}
\paragraph{Masked sample-level detection loss.}
A localization loss ensures that the detection of watermarked audio is done at the level of individual samples. 
For each time step $t$, we compute the binary cross entropy (BCE) between the detector's output $D(s)_t$ and the ground truth label (0 for non-watermarked, 1 for watermarked).
Overall, this reads:
\vspace{-0.2cm}
\begin{equation}\label{eq:loc_loss}
    \mathcal{L}_{loc}
    = \frac{1}{T} \sum_{t=1}^{T} \mathrm{BCE}(D(s')_t, y_t),
    \vspace{-0.3cm}
\end{equation}
where $s'$ might be $s$ or $s_w$, and where time step labels $y_t$ are set to 1 if they are watermarked, and 0 otherwise.

\subsection{Multi-bit watermarking}

We extend the method to support multi-bit watermarking, which allows for attribution of audio to a specific model version.
\emph{At generation}, we add a message processing layer in the middle of the generator.
It takes the activation map in $\R ^{h, t'}$ and a binary message $m\in \{0,1\} ^{b}$ and outputs a new activation map to be added to the original one.
We embed $m$ into $ e = \sum_{i=0..b-1}{E_{2i + m_i} \in \R^h}$, where $E\in \R^{2b, h}$ is a learnable embedding layer.
$e$ is then repeated $t$ times along the temporal axis to match the activation map size ($t,h$).
\emph{At detection}, we add $b$ linear layers at the very end of the detector. 
Each of them outputs a soft value for each bit of the message at the sample-level.
Therefore, the detector outputs a tensor of shape $\R^{t, 1+b}$ (1 for the detection, $b$ for the message).
\emph{At training}, we add a decoding loss $\mathcal{L}_{dec}$ to the localization loss $\mathcal{L}_{loc}$.
This loss $\mathcal{L}_{dec}$ averages the BCE between the original message and the detector's outputs over all parts where the watermark is present.

\subsection{Training details}
Our watermark generator and detector are trained on a 4.5K hours subset from the VoxPopuli~\cite{voxpopuli-WangRLWTHWPD20} dataset.
It is important to emphasize that the sole purpose of our generator is to generate imperceptible watermarks given an input audio; without the capability to produce or modify speech content.
We use a sampling rate of 16~kHz and one-second samples, so $T=16000$ in our training.
A full training requires 600k steps, with Adam, a learning rate of $10^{-4}$, and a batch size of 32.
For the drop augmentation, we use $k=5$ windows of $0.1$ sec. $h$ is set to 32, and the number of additional bits $b$ to 16 (note that $h$ needs to be higher than $b$, for example $h=8$ is enough in the zero-bit case). The perceptual losses are balanced and weighted as follows: $\lambda_{\ell_1} = 0.1$, $\lambda_{msspec} = 2.0$, $\lambda_{adv} = 4.0$, $\lambda_{loud} = 10.0$. The localization and watermarking losses are weighted by $\lambda_{loc} = 10.0$ and $\lambda_{dec} = 1.0$ respectively.

\subsection{Detection, localization and attribution}
At inference, we may use the generator and detector for:
\begin{itemize}[itemsep=1pt, topsep=1pt, leftmargin=8pt]
\item \emph{Detection}: To determine if the audio is watermarked or not. 
To achieve this, we use the average detector's output over the entire audio and flag it if the score exceeds a threshold (default: 0.5).
\item \emph{Localization}: To precisely identify where the watermark is present. We utilize the sample-wise detector's output and mark a time step as watermarked if the score surpasses a threshold (default: 0.5).
\item \emph{Attribution}: To identify the model version that produced the audio, enabling differentiation between users or APIs with a single detector. 
The detector's first output gives the detection score and the remaining $k$ outputs are used for attribution. 
This is done by computing the average message over detected samples and returning the identifier with the smallest Hamming distance.
\end{itemize}

\section{Audio/Speech Quality}
\label{sec:quality}

We first evaluate the quality of the watermarked audio using:
Scale Invariant Signal to Noise Ratio (SI-SNR): 
$\textrm{SI-SNR}(s, s_w) = 10 \log_{10} \left( \| \alpha s \|_2^2 / \| \alpha s - s_w \|_2^2 \right)$,
where $\alpha = \langle s, s_w \rangle / \| s \|_2^2$;
as well as PESQ~\cite{rix2001perceptual}, ViSQOL~\cite{hines2012visqol} and STOI~\cite{taal2010short} which are objective perceptual metrics measuring the quality of speech signals.

\autoref{tab:audio_quality} report these metrics.
\ours behaves differently than watermarking methods like \wavmark~\cite{chen2023wavmark} that try to minimize the SI-SNR.
In practice, high SI-SNR is indeed not necessarily correlated with good perceptual quality.
\ours is not optimized for SI-SNR but rather for perceptual quality of speech.
This is better captured by the other metrics (PESQ, STOI, ViSQOL), where \ours consistently achieves better performance.
Put differently, our goal is to hide as much watermark power as possible while keeping it perceptually indistinguishable from the original.
\autoref{fig:loc_quali} also visualizes how the watermark signal follows the shape of the speech waveform.

The metric used for our subjective evaluations is MUSHRA test~\cite{mushra}. The complete details about our full protocol can be found in the Appendix~\ref{app:MUSHRA}. In this study our samples got ratings very close to the ground truth samples that obtained an average score of $80.49$.

\begin{table}[t]
    \centering
    \caption{
        \textbf{Audio quality metrics}. 
        Compared to traditional watermarking methods that minimize the SNR like \wavmark, \ours achieves same or better perceptual quality.
    }\label{tab:audio_quality}
    \vspace{0.2cm}
    \resizebox{1.0\linewidth}{!}{
        \begin{tabular}{lccccc}
            \toprule
            \textbf{Methods} & \textbf{SI-SNR} & \textbf{PESQ} & \textbf{STOI} & \textbf{ViSQOL} & \textbf{MUSHRA}  \\
            \midrule
            \wavmark & \textbf{38.25} & 4.302 & 0.997 & 4.730 & 71.52 $\pm$ 7.18\\
            \ours & 26.00 & \textbf{4.470} & 0.997 & \textbf{4.829} &  \textbf{77.07} $\pm$ 6.35 \\
            \bottomrule
        \end{tabular}
    }
    \vspace{-8pt}
\end{table}

\section{Experiments and Evaluation}
\label{sec:exps}
This section evaluates the detection performance of passive classifiers, watermarking methods, and \ours, using True Positive Rate (TPR) and False Positive Rate (FPR) as key metrics for watermark detection. 
TPR measures correct identification of watermarked samples, while FPR indicates the rate of genuine audio clips falsely flagged.
In practical scenarios, minimizing FPR is crucial. For example, on a platform processing 1 billion samples daily, an FPR of $10^{-3}$ and a TPR of $0.5$ means that 1 million samples require manual review each day, yet only half of the watermarked samples are detected.

\subsection{Comparison with passive classifier}\label{sec:active-passive}

We first compare detection results on samples generated with Voicebox~\cite{le2023voicebox}.
We compare to the passive setup where a classifier is trained to discriminate between Voicebox-generated and real audios.
Following the approach in the Voicebox study, we evaluate 2,000 approximately 5-second samples from LibriSpeech, These samples have masked frames (90\%, 50\%, and 30\% of the phonemes) pre-Voicebox generation.
We evaluate on the same tasks, \ie distinguishing between original and generated, or between original and re-synthesized (created by extracting the Mel spectrogram from original audio and then vocoding it with the HiFi-GAN vocoder).

Both active and passive setups achieve perfect classification in the case when trained to distinguish between natural and Voicebox.
Conversely, the second part of Tab.~\ref{tab:voicebox} highlights a significant drop in performance when the classifier is trained to differentiate between Voicebox and re-synthesized.
It suggests that the classifier is detecting vocoder artifacts, since the re-synthesized samples are sometimes wrongly flagged.
The classification performance quickly decreases as the quality of the AI-generated sample increases (when the input is less masked).
On the other hand, our proactive detection does not rely on model-specific artifacts but on the watermark presence. %
This allows for perfect detection over all the audio clips. %

\begin{table}[t]
    \centering
    \caption{
        \textbf{Comparison with Voicebox binary classifier.} 
        Percentage refers to the fraction of masked input frames.
    }
    \label{tab:voicebox}
    \vspace{4pt}
    \resizebox{0.95\linewidth}{!}{
        \begin{tabular}{r *{3}{c}  *{3}{c}  *{3}{c} }
            \toprule
            & \multicolumn{3}{c}{\textbf{\ours (Ours)}} & \multicolumn{3}{c}{\textbf{Voicebox Classif.}} \\
            \cmidrule(rr){2-4} \cmidrule(rr){5-7}
            \textbf{\% Mask} & Acc. & TPR & FPR & Acc. & TPR & FPR \\
            \midrule
            \multicolumn{7}{l}{\emph{Original audio vs AI-generated audio}} \\
            30\% & 1.0 & 1.0 & 0.0 & 1.0 & 1.0 & 0.0 \\
            50\% & 1.0 & 1.0 & 0.0 & 1.0 & 1.0 & 0.0 \\
            90\% & 1.0 & 1.0 & 0.0 & 1.0 & 1.0 & 0.0 \\
            \midrule
            \multicolumn{7}{l}{\emph{Re-synthesized audio vs AI-generated audio}} \\
            30\% & \textbf{1.0} & \textbf{1.0} & \textbf{0.0} & 0.704 & 0.680 & 0.194 \\
            50\% & \textbf{1.0} & \textbf{1.0} & \textbf{0.0} & 0.809 & 0.831 & 0.170 \\
            90\% & \textbf{1.0} & \textbf{1.0} & \textbf{0.0} & 0.907 & 0.942 & 0.112 \\
            \bottomrule
        \end{tabular}
    }
    \vspace{-0.2cm}
\end{table}

\subsection{Comparison with watermarking}
\newcommand{\aux}[1]{{\scriptsize{\textcolor{gray}{#1}}}}
\begin{table}[t]
    \centering
    \caption{
        \textbf{Detection results} for different edits applied before detection. 
        Acc. ({\aux{TPR/FPR}}) is the accuracy (and TPR/FPR) obtained for the threshold that gives best accuracy on a balanced set of augmented samples.
        AUC is the area under the ROC curve.
    }
    \label{tab:wm_robustness}
    \vspace{4pt}
    \resizebox{1.0\linewidth}{!}{
        \begin{tabular}{l *{2}{l}  *{2}{l}}
        \toprule
        & \multicolumn{2}{l}{\textbf{\ours (Ours)}} & \multicolumn{2}{l}{\textbf{\wavmark}} \\
        \cmidrule(rr){2-3} \cmidrule(rr){4-5}
        \multicolumn{1}{c}{Edit} & Acc. \aux{TPR/FPR} & AUC & Acc. \aux{TPR/FPR}  & AUC \\
        \cmidrule(rr){1-1} \cmidrule(rr){2-3} \cmidrule(rr){4-5}
        None & 1.00 \aux{1.00/0.00} & 1.00 & 1.00 \aux{1.00/0.00} & 1.00 \\
        Bandpass & 1.00 \aux{1.00/0.00} & 1.00 & 1.00 \aux{1.00/0.00} & 1.00 \\
        Highpass  &  0.61 \aux{0.82/0.60} & 0.61 & \bf 1.00 \aux{1.00/0.00} & \bf 1.00 \\
        Lowpass & \bf 0.99 \aux{0.99/0.00} & \bf 0.99 & 0.50 \aux{1.00/1.00} & 0.50 \\
        Boost & 1.00 \aux{1.00/0.00} & 1.00 & 1.00 \aux{1.00/0.00} & 1.00 \\
        Duck & 1.00 \aux{1.00/0.00} &  1.00 & 1.00 \aux{1.00/0.00} & 1.00 \\
        Echo & \bf 1.00 \aux{1.00/0.00} & \bf 1.00 & 0.93 \aux{0.89/0.03} & 0.98 \\
        Pink & \bf 1.00 \aux{1.00/0.00} & \bf 1.00 & 0.88 \aux{0.81/0.05} & 0.93 \\
        White & \bf 0.91 \aux{0.86/0.04} & \bf 0.95 & 0.50 \aux{0.54/0.54} & 0.50 \\
        Fast (1.25x) & \bf 0.99 \aux{0.99/0.00} & \bf 1.00 & 0.50 \aux{0.01/0.00} & 0.15 \\
        Smooth & \bf 0.99 \aux{0.99/0.00} &  1.00   & 0.94 \aux{0.93/0.04} & 0.98 \\
        Resample & 1.00 \aux{1.00/0.00} &  1.00 & 1.00 \aux{1.00/0.00} & 1.00 \\
        AAC & 1.00 \aux{1.00/0.00} &  1.00 & 1.00 \aux{1.00/0.00} & 1.00 \\
        MP3 & \bf 1.00 \aux{1.00/0.00} & \bf 1.00 & 1.00 \aux{0.99/0.00} & 0.99 \\
        EnCodec & \bf  0.98 \aux{0.98/0.01} & \bf 1.00 & 0.51 \aux{0.52/0.50} & 0.50 \\
        \midrule
        Average & \bf 0.96 \aux{0.98/0.04} & \bf 0.97 & 0.85 \aux{0.85/0.14} & 0.84 \\
        \bottomrule
        \end{tabular}
    }
    \vspace{-0.2cm}
\end{table}

We evaluate the robustness of the detection on a wide range of audio editing operations: 
time modification (faster, resample), 
filtering (bandpass, highpass, lowpass), 
audio effects (echo, boost audio, duck audio), 
noise (pink noise, random noise),
and compression (MP3, AAC, EnCodec).
These attacks cover a wide range of transformations that are commonly used in audio editing software.
For all edits except EnCodec compression, evaluation with parameters in the training range would be perfect. In order to show generalization, we chose stronger parameter to the attacks than those used during training (details in App.~\ref{app:augmentations}). 

Detection is done on 10k ten-seconds audios from our VoxPopuli validation set.
For each edit, we first build a balanced dataset made of the 10k watermarked/ 10k non-watermarked edited audio clips.
We quantify the performance by adjusting the threshold of the detection score, selecting the value that maximizes accuracy (we provide corresponding TPR and FPR at this threshold).
The ROC AUC (Area Under the Curve of the Receiver Operating Characteristics) gives a global measure of performance over all threshold levels, and captures the TPR/FPR trade-off.
To adapt data-hiding methods (\eg \wavmark) for proactive detection, we embed a binary message (chosen randomly beforehand) in the generated speech before release. The detection score is then computed as the Hamming distance between the original message and the one extracted from the scrutinized audio. 

We observe in Tab.~\ref{tab:wm_robustness} that \ours is overall more robust, with an average AUC of 0.97 vs. 0.84 for \wavmark.
The performance for lowpass and highpass filters indicates that \ours embeds watermarks neither in the low nor in the high frequencies (\wavmark focuses on high frequencies).
We give results on more augmentations in App.~\ref{app:robustness}.

\paragraph{Generalization.} 
\label{sec:generalization}
We evaluate how \ours generalizes on various domains and languages. 
Specifically, we use the datasets ASVspoof~\cite{liu2023asvspoof} and FakeAVCeleb~\cite{khalid2021fakeavceleb}. Additionally, we translate speech samples from a subset of the Expresso dataset~\cite{nguyen2023expresso} (studio-quality recordings) using the SeamlessExpressive translation model~\cite{seamless2023}.
We select four target languages: Mandarin Chinese (CMN), French (FR), Italian (IT), and Spanish (SP). 
We also evaluate on non-speech AI-generated audios: music from MusicGen~\cite{copet2023simple} and environmental sounds from AudioGen~\cite{kreuk2023audiogen}. 
Results are very similar to our in-domain test set and can be found in App.~\ref{app:ood}.

\subsection{Localization}

\begin{figure}[b!]
    \vspace{-0.2cm}
    \centering
    \includegraphics[width=0.85\linewidth, clip, trim={0 0 0 0}]{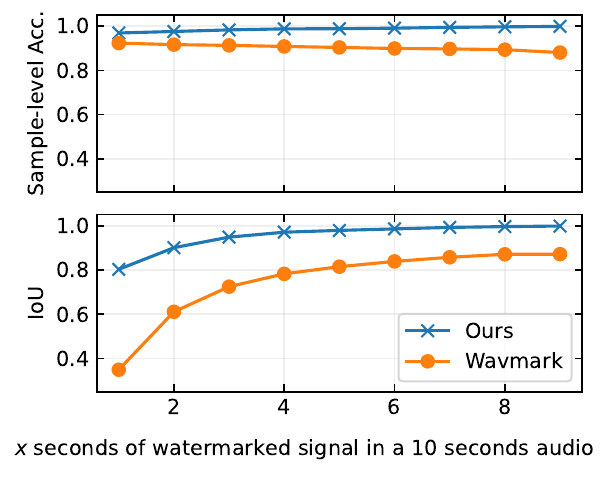}
    \vspace{-0.3cm}
    \caption{\textbf{Localization results} across different durations of watermarked audio signals in terms of Sample-Level Accuracy and Intersection Over Union (IoU) metrics ($\uparrow$ is better).}
    \label{fig:loc_quantitative}
\end{figure}

We evaluate localization with the sample-level detection accuracy, \ie the proportion of correctly labeled samples, and the Intersection over Union (IoU).
The latter is defined as the intersection between the predicted and the ground truth detection masks (1 when watermarked, 0 otherwise), divided by their union.
IoU is a more relevant evaluation of the localization of short watermarks in a longer audio.

This evaluation is carried out on the same audio clips as for detection.
For each one of them, we watermark a randomly placed segment of varying length.
Localization with \wavmark is a brute-force detection: a window of 1s slides over the 10s of speech with the default shift value of 0.05s.
The Hammning distance between the 16 pattern bits is used as the detection score.
Whenever a window triggers a positive, we label its 16k samples as watermarked in the detection mask in $\{0,1\}^t$.

\autoref{fig:loc_quantitative} plots the sample-level accuracy and IoU
for different proportions of watermarked speech in the audio clip.
\ours achieves an IoU of 0.99 when just one second of speech is AI-manipulated, compared to \wavmark's 0.35.
Moreover, \ours allows for precise detection of minor audio alterations: it can pinpoint AI-generated segments in audio down to the sample level (usually 1/16k sec), while the concurrent \wavmark only provides one-second resolution and therefore lags behind in terms of IoU. 
This is especially relevant for speech samples, where a simple word modification may greatly change meaning.

\subsection{Attribution}

Given an audio clip, the objective is now to find if any of $N$ versions of our model generated it (detection), and if so, which one (identification). 
For evaluation, we create $N'=100$ random 16-bits messages and use them to watermark 1k audio clips, each consisting of 5 seconds of speech (not 10s to reduce compute needs). 
This results in a total of 100k audios. 
For \wavmark, the first 16 bits (/32) are fixed and the detection score is the number of well decoded pattern bits, while the second half of the payload hides the model version.
An audio clip is flagged if the average output of the detector exceeds a threshold, 
corresponding to FPR=$10^{-3}$.
Next, we calculate the Hamming distance between the decoded watermark and all $N$ original messages. 
The message with the smallest Hamming distance is selected.
It's worth noting that we can simulate $N>N'$ models by adding extra messages. 
This may represent versions that have not generated any sample.

False Attribution Rate (FAR) is the fraction of wrong attribution \emph{among the detected audios} while the attribution accuracy is the  
proportion of detections followed by a correct attributions \emph{over all audios}. \ours has a higher FAR but overall gives a better accuracy, which is what ultimately matters.
In summary, decoupling detection and attribution achieves better detection rate and makes the global accuracy better, at the cost of occasional false attributions.

\begin{table}[t]
    \centering
    \caption{
        \textbf{Attribution results}.
        We report the accuracy of the attribution (Acc.) and false attribution rate (FAR). 
        Detection is done at FPR=$10^{-3}$ and attribution matches the decoded message to one of $N$ versions.
        We report averaged results over the edits of Tab.~\ref{tab:wm_robustness}.
    }\label{tab:attribution}
    \vspace{0.1cm}
    \resizebox{0.99\linewidth}{!}{
        \begin{tabular}{cr *{5}{c}}
            \toprule
            & N & $1$ & $10$ & $10^2$ & $10^3$ & $10^4$ \\ \midrule
    \multirow{2}{*}{FAR (\%) $\downarrow$} & \wavmark      & 0.0 & \textbf{0.20} & \textbf{0.98} & \textbf{1.87} & \textbf{4.02} \\
            & \ours   & 0.0 & 2.52 & 6.83 & 8.96 & 11.84 \\ \midrule
    \multirow{2}{*}{\shortstack{Acc. (\%) $\uparrow$}} & \wavmark      & 58.4 & 58.2 & 57.4 & 56.6 & 54.4 \\
            & \ours  & \textbf{68.2} & \textbf{65.4} & \textbf{61.4} & \textbf{59.3} & \textbf{56.4} \\ 
            \bottomrule
        \end{tabular}
    }
    \vspace{-0.2cm}
\end{table}

\begin{figure}[b]
    \centering
    \includegraphics[width=0.95\linewidth, clip, trim={0 0 0 0}]{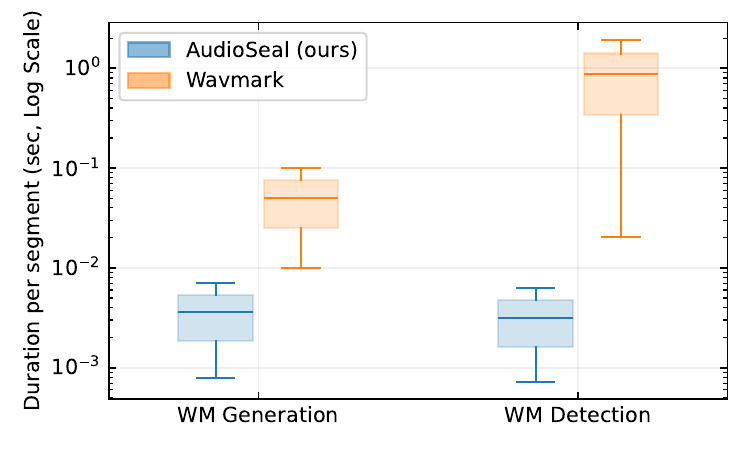}
    \vspace{-8pt}
    \caption{\textbf{Mean runtime} ($\downarrow$ is better).
    \ours is one order of magnitude faster for watermark generation and two orders of magnitude faster for watermark detection for the same audio input. See Appendix \ref{app:speed} for full comparison. 
    }
    \label{fig:efficiency}
\end{figure}

\subsection{Efficiency Analysis}
\label{sec:speed}
To highlight the efficiency of \ours, we conduct a performance analysis and compare it with \wavmark. 
We apply the watermark generator and detector of both models on a dataset of 500 audio segments ranging in length from 1 to 10 seconds, using a single Nvidia Quadro GP100 GPU. 
The results are displayed in Fig.~\ref{fig:efficiency} and Tab.~\ref{tab:speed}. 
In terms of generation, \ours is 14x faster than \wavmark. 
For detection, \ours outperforms \wavmark with two orders of magnitude faster performance on average, notably 485x faster in scenarios where there is no watermark (Tab.~\ref{tab:speed}). 
This remarkable speed increase is due to our model's unique localized watermark design, which bypasses the need for watermark synchronization (recall that \wavmark relies on 20 pass forwards for a one-second snippet).
\ours's detector provides detection logits for each input sample directly with only one pass to the detector, significantly enhancing the detection's computational efficiency.
This makes our system highly suitable for real-time and large-scale applications.

\section{Adversarial Watermark Removal}
\label{sec:attacks}

\begin{figure}[b]
    \centering
    \includegraphics[width=0.95\linewidth, clip, trim={0 0.3cm 0 0}]{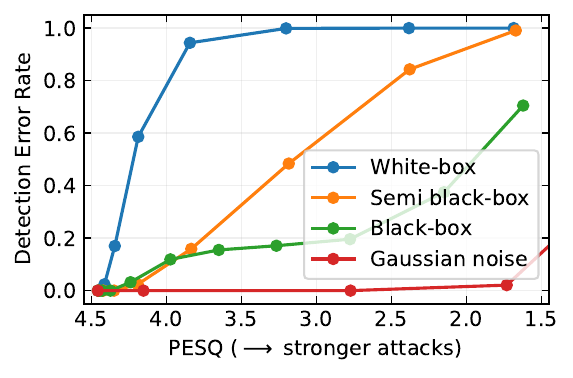}
    \vspace{-0.2cm}
    \caption{
    \textbf{Watermark-removal attacks.} 
    PESQ is measured between attacked audios and genuine ones (PESQ $<4$ strongly degrades the audio quality).
    The more knowledge the attacker has over the watermarking algorithm, the better the attack is.
    }
    \label{fig:attacks}
\end{figure}

We now examine more damaging deliberate attacks, where attackers might either ``forge'' the watermark by adding it to authentic samples (to overwhelm detection systems) or ``remove'' it to avoid detection. 
Our findings suggest that in order to maintain the effectiveness of watermarking against such adversaries, the code for training watermarking models and the awareness that published audios are watermarked can be made public. 
However, the detector's weights should be kept confidential.

We focus on watermark-removal attacks and consider three types of attacks depending on the adversary's knowledge:
\begin{itemize}[leftmargin=*, topsep=2pt, itemsep=2pt]
    \item \textit{White-box}: 
    the adversary has access to the detector (\eg because of a leak), and performs a gradient-based adversarial attack against it.
    The optimization objective is to minimize the detector's output.
    \item \textit{Semi black-box}: 
    the adversary does not have access to any weights, but is able to re-train generator/detector pairs with the same architectures on the same dataset.
    They perform the same gradient-based attack as before, but using the new detector as proxy for the original one.
    \item \textit{Black-box}: 
    the adversary does not have any knowledge on the watermarking algorithm being used, but has access to an API that produces watermarked samples, and to negative speech samples from any public dataset.
    They first collect samples and train a classifier to discriminate between watermarked and not-watermarked.
    They attack this classifier as if it were the true detector.
\end{itemize}

For every scenario, we watermark 1k samples of 5 seconds, then attack them.
The gradient-based attack optimizes an adversarial noise added to the audio, with 100 steps of Adam.
During the optimization, we control the norm of the noise to trade off attack strength and audio quality.
When training the classifier for the black-box attack, we use 80k/80k watermarked/genuine samples of 8 seconds and make sure the classifier has 100\% detection accuracy on the validation set.
More details in App.~\ref{app:attacks}.

\autoref{fig:attacks} contrasts various attacks at different intensities, using Gaussian noise as a reference.
The white-box attack is by far the most effective one, increasing the detection error by around 80\%, while maintaining high audio quality (PESQ $>4$).
Other attacks are less effective, requiring significant audio quality degradation to achieve $50\%$ increase the detection error, though they are still more effective than random noise addition.
In summary, the more is disclosed about the watermarking algorithm, the more vulnerable it is. 
The effectiveness of these attacks is limited as long as the detector remains confidential.

\section{Conclusion}

In this paper, we introduced \ours, a proactive method for the detection, localization, and attribution of AI-generated speech. 
\ours revamps the design of audio watermarking to be specific to localized detection rather than data hiding. 
It is based on a generator/detector architecture that can generate and extract watermarks at the audio sample level. 
This removes the dependency on slow brute force algorithms, traditionally used to encode and decode audio watermarks.
The networks are jointly trained through a novel loudness loss, differentiable augmentations and masked sample level detection losses. 
As a result, \ours achieves state-of-the-art robustness to various audio editing techniques, very high precision in localization, and orders of magnitude faster runtime than methods relying on synchronization.    
Through an empirical analysis of possible adversarial attacks, we conclude that for watermarking to still be an effective mitigation, the detector's weights have to be kept private -- otherwise adversarial attacks might be easily forged. 
A key advantage of \ours is its practical applicability. 
It stands as a ready-to-deploy solution for watermarking in voice synthesis APIs. 
This is pivotal for large-scale content provenance on social media and for detecting and eliminating incidents, enabling swift action on instances like the US voters' deepfake case~\cite{murphy2024biden} long before they spread.

\section*{Impact Statement}
This research aims to improve transparency and traceability in AI-generated content, but watermarking in general can have a set of potential misuses such as government surveillance of dissidents or corporate identification of whistle blowers. 
Additionally, the watermarking technology might be misused to enforce copyright on user-generated content, and its ability to detect AI-generated audio could increase skepticism about digital communication authenticity, potentially undermining trust in digital media and AI. 
However, despite these risks, ensuring the detectability of AI-generated content is important, along with advocating for robust security measures and legal frameworks to govern the technology's use.

\bibliographystyle{icml_conference}
\bibliography{references}

\clearpage
\appendix

\section{Extended related work\label{app:related}}

\paragraph{Zero-shot TTS and vocal style preservation.}
There has been an emergence of models that imitate or preserve vocal style using only a small amount of data. 
One key example is zero-shot text-to-speech (TTS) models.
These models create speech in vocal styles they haven't been specifically trained on. 
For instance, models like VALL-E~\cite{wang2023neural}, YourTTS~\cite{casanova2022yourtts}, NaturalSpeech2 \cite{shen_naturalspeech2} synthesize high-quality personalized speech with only a 3-second recording. 
On top, zero-shot TTS models like Voicebox ~\cite{le2023voicebox}, A$^{3}$T~\cite{BaiZCML022_A3T} and Audiobox~\cite{hsu2023audiobox}, with their non-autoregressive inference, perform tasks such as text-guided speech infilling, where the goal is to generate masked speech given its surrounding audio and text transcript. 
It makes them a powerful tool for speech manipulation.
In the context of speech machine translation, SeamlessExpressive~\cite{seamless2023} is a model that not only translates speech, but also retains the speaker's unique vocal style and emotional inflections, thereby broadening the capabilities of such systems.

\vspace*{-4pt}\paragraph{Audio generation and compression.}
Early models are  autoregressive like WaveNet~\cite{wavnet}, with dilated convolutions and waveform reconstruction as objective.
Subsequent approaches explore different audio losses, such as scale-invariant signal-to-noise ratio (SI-SNR)~\cite{convtasnet} or Mel spectrogram distance~\cite{defossez2020real}. 
None of these objectives are deemed ideal for audio quality, leading to the adoption of adversarial models in HiFi-GAN~\cite{hifigan} or MelGAN~\cite{melgan}. 
Our training objectives and architectures are inspired by more recent neural audio compression models~\cite{defossez2022high, dac, soundstream}, that focus on high-quality waveform generation and integrate a combination of these diverse objectives in their training processes.

\paragraph{Synchronization and Detection speed.} 
To accurately extract watermarks, synchronization between the encoder and decoder is crucial. 
However, this can be disrupted by desynchronization attacks such as time and pitch scaling. 
To address this issue, various techniques have been developed. 
One approach is block repetition, which repeats the watermark signal along both the time and frequency domains~\cite{blockrep2-KirovskiM03, blockrep1-Kirovski2003}. 
Another method involves implanting synchronization bits into the watermarked signal~\cite{patchwork-sync-XiangNGZN14}. 
During decoding, these synchronization bits serve to improve synchronization and mitigate the effects of de-synchronization attacks. 
Detection of those synchronization bits for watermark detection usually involves exhaustive search using brute force algorithms, which significantly slows down decoding time.

\begin{table*}[t]
    \centering
    \caption{
        The average runtime (ms) per sample of our proposed \ours model against the state-of-the-art Wavmark\cite{chen2023wavmark} method. Our experiments were conducted on a dataset of audio segments spanning 1 sec to 10 secs, using a single Nvidia Quadro GP100 GPU. The results, displayed in the table, demonstrate substantial speed enhancements for both Watermark Generation and Detection with and without the presence of a watermark. Notably, for watermark detection, AudioSeal is \textbf{485$\times$ faster} than Wavmark during the absence of a watermark, more details in section \ref{sec:speed}. 
    }
    \footnotesize
    \label{tab:speed}
    \vspace{4pt}
    \begin{tabular}{llll}
    \toprule
               Model & Watermarked &     \textbf{Detection ms (speedup)} &   \textbf{Generation ms (speedup)} \\
    \midrule
             Wavmark &          No & 1710.70 $\pm$ 1314.02 &    -- \\
             AudioSeal (ours) &          No &       \textbf{3.25 $\pm$ 1.99} \;\; (\textbf{485$\times$}) &    -- \\
    \midrule
             Wavmark &         Yes &    106.21 $\pm$ 66.95 & 104.58 $\pm$ 65.66 \\
    AudioSeal (ours) &         Yes &       \textbf{3.30} $\pm$ \textbf{2.03} \;\; (\textbf{35$\times$}) &    \textbf{7.41} $\pm$ \textbf{4.52} \;\; (\textbf{14} $\times$) \\
    
    \bottomrule 
    \end{tabular}
\end{table*}

\section{False Positive Rates - Theory and Practice}\label{app:fpr}

\paragraph{Theoretical FPR.}

\begin{figure}[b]
    \centering
    \includegraphics[width=0.99\linewidth, clip, trim={0.1in 0 0.1in 0}]{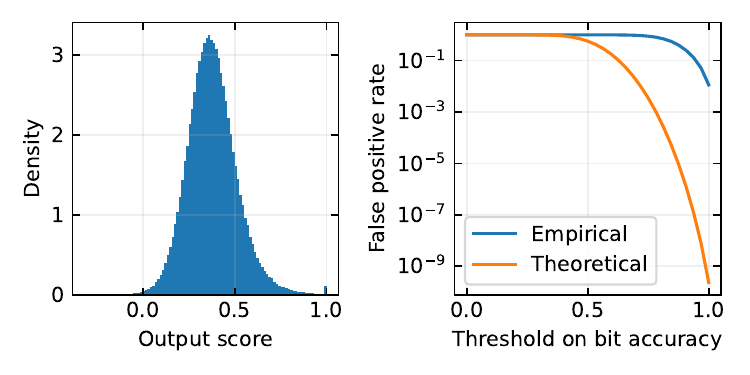}
    \vspace{-0.3cm}
    \caption{
        (Left) Histogram of scores output by WavMark's extractor on 10k genuine samples. 
        (Right) Empirical and theoretical FPR when the chosen hidden message is all 0.
    }
    \label{fig:app_fpr_wavmark}
\end{figure}

When doing multi-bit watermarking, previous works~\cite{yu2021artificial, kim2023wouaf, fernandez2023stable, chen2023wavmark} usually extract the message $m'$ from the content $x$ and compare it to the original binary signature $m\in \{ 0,1 \}^{k}$ embedded in the speech sample.
The detection test relies on the number of matching bits $M(m,m')$:
\begin{equation} 
    \text{if } M\left(m,m'\right) \geq \tau \,\,\textrm{ where }\,\, \tau\in 
\{0,\ldots,k\},
\end{equation}
then the audio is flagged.
This provides theoretical guarantees over the false positive rates.

Formally, the statistical hypotheses are $H_1$: ``The audio signal $x$ is watermarked'', and the null hypothesis $H_0$: ``The audio signal $x$ is genuine''.
Under $H_0$ (\ie, for unmarked audio), if the bits $m'_1, \ldots, m'_k$ are independent and identically distributed (i.i.d.) Bernoulli random variables with parameter $0.5$, then  $M(m, m')$ follows a binomial distribution with parameters ($k$, $0.5$).
The False Positive Rate (FPR) is defined as the probability that $M(m, m')$ exceeds a given threshold $\tau$. 
A closed-form expression can be given using the regularized incomplete beta function $I_x(a;b)$ (linked to the CDF of the binomial distribution):
\begin{align}\label{eq:p-value}
    \text{FPR}(\tau) & = \mathbb{P}\left(M \geq \tau | H_0\right) = I_{1/2}(\tau, k - \tau +1).
\end{align}

\paragraph{Empirical study.}
We empirically study the FPR of WavMark-based detection on our validation dataset.
We use the same parameters as in the original paper, \ie $k=32$-bits are extracted from 1s speech samples.
We first extract the soft bits (before thresholding) from 10k genuine samples and plot the histogram of the scores in Fig.~\ref{fig:app_fpr_wavmark} (left).
We should observe a Gaussian distribution with mean $0.5$, while empirically the scores are centered around $0.38$. 
This makes the decision heavily biased towards bit 0 on genuine samples.
It is therefore impossible to theoretically set the FPR since this would largely underestimate the actual one.
For instance, Figure~\ref{fig:app_fpr_wavmark} (right) shows the theoretical and empirical FPR for different values of $\tau$ when the chosen hidden message is full 0.
Put differently, the argument that says that hiding bits allows for theoretical guarantees over the detection rates is not valid in practice.

\section{Additional Experimental Results}
\label{appendix:additional_exp}

\subsection{Computational efficiency}
\label{app:speed}

\begin{figure}[h]
    \centering
    \vspace{-0.4cm}
    \includegraphics[width=\linewidth, clip, trim={0 0 0 0}]{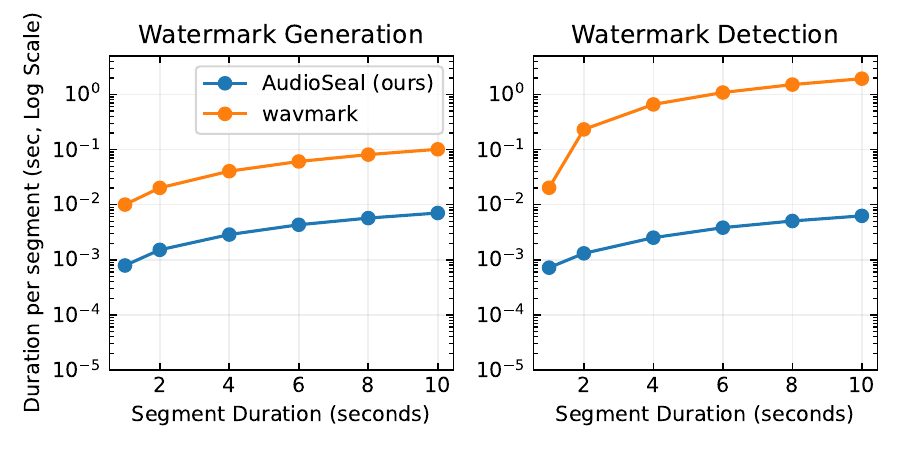}
    \vspace{-0.8cm}
    \caption{\textbf{Mean runtime} ($\downarrow$ is better) of \ours versus \wavmark. 
    AudioSeal is one order of magnitude faster for watermark generation andtwo orders of magnitude faster for watermark detection for the same audio input, signifying a considerable enhancement in real-time audio watermarking efficiency.
    }
    \label{fig:efficiency_vs_time}
\end{figure}

We show in \autoref{fig:efficiency_vs_time} the mean runtime of the detection and generation depending on the audio duration.
Corresponding numbers are given in \autoref{tab:speed}.

\subsection{Another architecture}
\label{app:other-arch}

Our architecture relies on the SOTA compression method EnCodec. 
However, to further validate our approach, we conduct an ablation study using a different architecture DPRNN~\cite{luo2020dual}. 
The results are presented in Tab.~\ref{tab:dprnn}.
They show that the performance of AudioSeal is consistent across different architectures, with similar performances using the much slower and heavier architecture from \citet{luo2020dual}. 
This indicates that model capacity is not a limiting factor for \ours.

\begin{table}[h]
\vspace{-0.2cm}
\centering
\caption{
    Results of AudioSeal with different architectures for the generator and detector.
    The IoU is computed for 1s of watermark in 10s audios (corresponding to the leftmost point in Fig.~\ref{fig:loc_quantitative}).
}\label{tab:dprnn}
\vspace{0.1cm}
\footnotesize
\begin{tabular}{lccccc}
    \toprule
    Method  & SISNR & STOI & PESQ & Acc. & IoU \\
    \midrule
    EnCodec & 26.00 & 0.997 & 4.470 & 1.00 & 0.802 \\
    DPRNN   & 26.7 & 0.996 & 4.421 & 1.00 & 0.796 \\
    \bottomrule
\end{tabular}
\vspace{-0.2cm}
\end{table}

\subsection{Audio mixing}

We hereby evaluate the scenario where two watermarked signals (e.g., vocal and instrumental) are mixed together. 
To explore this, we conducted experiments using a non-vocal music dataset. 
In these experiments, we normalized and summed the loudness of watermarked speech and music segments. 
The results are detailed Tab.~\ref{tab:mixed_signals}.

\begin{table}[h]
    \centering
    \caption{
        Detection results for watermarked speech and music mixed signals.
        \cmark and \xmark indicate the presence of the watermark.
    }
    \label{tab:mixed_signals}
    \vspace{0.1cm}
    \footnotesize
    \begin{tabular}{cccc}
    \toprule
    Speech & BG Music & Acc. \aux{FPR / TPR} & AUC \\
    \midrule
    \cmark & \cmark & 0.9996 \aux{0.0003 / 0.9996} & 0.9999 \\
    \cmark & \xmark & 0.9787 \aux{0.0310 / 0.9883} & 0.9961 \\
    \bottomrule
    \end{tabular}
    \vspace{-0.2cm}
\end{table}

\subsection{Out of domain (OOD) evaluations}\label{app:ood}

\begin{table*}[h]
    \caption{
    Evaluation of \ours Generalization across domains and languages. Namely, translations of speech samples from the Expresso dataset~\cite{nguyen2023expresso} to four target languages: Mandarin Chinese (CMN), French (FR), Italian (IT), and Spanish (SP), using the SeamlessExpressive model~\cite{seamless2023}. Music from MusicGen~\cite{copet2023simple} and environmental sounds from AudioGen~\cite{kreuk2023audiogen}. 
    \vspace{0.1cm}
    }
    \label{tab:ood_data}
    \resizebox{\textwidth}{!}{
        \begin{tabular}{l|cccccc|cc}
        \toprule
        Aug & Seamless (Cmn) & Seamless (Spa) & Seamless (Fra) & Seamless(Ita) & Seamless (Deu) & Voicebox (Eng) & AudioGen & MusicGen  \\
        \midrule
        None         & 1.00           & 1.00            & 1.00            & 1.00           & 1.00            &   1.00   &   1.00   &   1.00\\
        \midrule
        Bandpass   & 1.00 &   1.00  &   1.00 & 1.00 &   1.00  &  1.00 &   1.00   &   1.00   \\
            Highpass  & 0.71 &  0.68  &   0.70  & 0.70 &  0.70  &  0.64   &  0.52 &  0.52     \\
        Lowpass    & 1.00 &  0.99 &  1.00 & 1.00 & 1.00 &  1.00  &  1.00  &  1.00  \\
            Boost     & 1.00  &  1.00  &  1.00 & 1.00  &  1.00 &  1.00 &     1.00     &      1.00   \\
            Duck      & 1.00  &  1.00  &  1.00 &1.00 &  1.00 &   1.00  &   1.00   &     1.00    \\
            Echo   & 1.00  &  1.00 &  1.00 &1.00 & 1.00 &  1.00  &      1.00    &     1.00     \\
            Pink   & 0.99 &   1.00  & 0.99 & 1.00 &   0.99 &  1.00 &      1.00    &    1.00  \\
            White  & 1.00 &    1.00  & 1.00 & 1.00 &  1.00 & 1.00   &  1.00   &  1.00   \\
        Fast (x1.25)  & 0.97 & 0.98 & 0.99  & 0.98 & 0.99 & 0.98 & 0.87 & 0.87 \\
            Smooth    &  0.96  &  0.99  &   0.99  &    0.99    &      0.99          & 0.99 &  0.98  &    0.98   \\
            Resample  & 1.00 &  1.00 &  1.00 & 1.00 &    1.00  & 1.00 &   1.00    &   1.00   \\
                AAC & 0.99 &  0.99  &  0.99  & 0.99 &  0.99  &   0.97  &  0.99   &     0.98    \\
                MP3 & 0.99 &  0.99   &  0.99 & 0.99 & 0.99  &    0.97  &  0.99    & 1.00  \\
            Encodec   & 0.97 &  0.98   &  0.99 & 0.99 & 0.98 &   0.96     &  0.95    & 0.95   \\
            \midrule
            Average   &  0.97 &  0.97  & 0.98  & 0.98 & 0.98 & 0.97 & 0.95 & 0.95  \\
            \bottomrule
        \end{tabular}
    }
\end{table*}

As previously outlined in Sec.~\ref{sec:generalization}, we tested \ours on the outputs of various voice cloning models and other audio modalities. 
We employed the same set of augmentations and observed very similar results, as demonstrated in Tab.~\ref{tab:ood_data}.
Interestingly, even though we did not train our model on AI-generated speech, we noticed an improvement in performance compared to our test data. 
No sample was misclassified among the 10k samples that comprised each of our out-of-distribution (OOD) datasets.
We also provide the other perceptual metrics results on OOD data in Tab.~\ref{tab:ood_metrics}.

We also evaluated \ours on three additional datasets containing real human speech: AudioSet~\cite{gemmeke2017audio}, ASVspoof~\cite{liu2023asvspoof}, and FakeAVCeleb~\cite{khalid2021fakeavceleb}.
Again, we observed similar performance, as shown in Tab.~\ref{tab:other_datasets}.

\begin{table}[h]
\centering
\caption{Audio quality and intelligibility evaluations on AI generated speech data from various models and languages.}
\label{tab:ood_metrics}
\vspace{0.1cm}
\footnotesize
\renewcommand{\arraystretch}{1.2}
\resizebox{1.0\linewidth}{!}{
\begin{tabular}{cccccc}
    \toprule
    Model & Dataset & SISNR & PESQ & STOI & VISQOL \\
    \midrule
    \multirow{3}{*}{\scriptsize \rotatebox[origin=c]{90}{\ours}} & Seam. (Deu)       & 23.35 & 4.244 & 0.999 & 4.688 \\
     & Seam. (Fr)        & 24.02 & 4.199 & 0.998 & 4.669 \\
     & Voicebox             & 25.23 & 4.449 & 0.998 & 4.800 \\
     \midrule
    \multirow{3}{*}{\scriptsize \rotatebox[origin=c]{90}{\wavmark}} & Seam. (Deu)    & 38.93 & 3.982 & 0.999 & 4.515 \\
     & Seam. (Fr)     & 39.06 & 3.959 & 0.999 & 4.506 \\
     & Voicebox          & 39.63 & 4.211 & 0.998 & 4.695 \\
    \bottomrule
\end{tabular}
\vspace{-0.2cm}
}
\end{table}

\begin{table}[h]
    \centering
    \caption{Evaluation of the detection performances on different datasets. AudioSet is an environmental sounds dataset while ASVspoof~\cite{liu2023asvspoof} and FakeAVCeleb~\cite{khalid2021fakeavceleb} are deep-fake detection datasets.}
    \vspace*{0.2cm}
    \label{tab:other_datasets}
    \footnotesize
    \begin{tabular}{l *{2}{l}}
        \toprule
        Dataset & Acc. \aux{TPR/FPR} & AUC \\
        \midrule
        Audioset & 0.9992 \aux{0.9996/0.0011} & 1.0 \\
        ASVspoof & 1.0 \aux{1.0/0.0} & 1.0 \\
        FakeAVCeleb & 1.0 \aux{1.0/0.0} & 1.0 \\
        \bottomrule
    \end{tabular}
    \vspace{-0.2cm}
\end{table}

\begin{figure*}[h]
    \centering
    \includegraphics[width=0.9\textwidth]{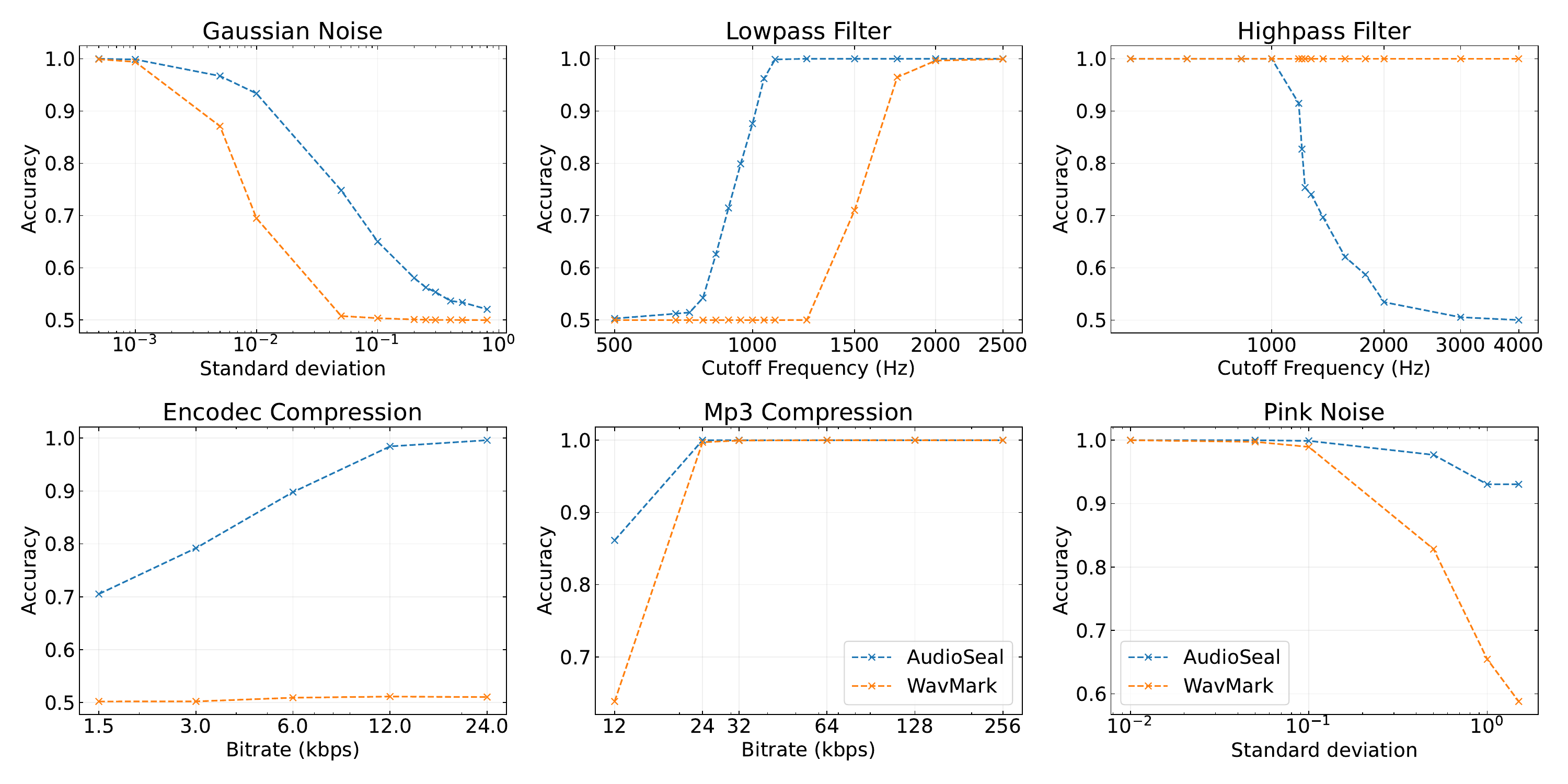}
    \vspace{-0.4cm}
    \caption{
        Accuracy of the detector on augmented samples with respect to the strength of the augmentation.\label{fig:app_augmentation_curves}
    }
    \label{fig:enter-label}
\end{figure*}

\subsection{Robustness results}\label{app:robustness}

We plot the detection accuracy against the strength of multiple augmentations in Fig.~\ref{fig:app_augmentation_curves}. 
\ours outperforms \wavmark for most augmentations at the same strength.
However, for highpass filters above our training range (500Hz) \wavmark has a much better detection accuracy.
Our system's TF-loudness loss embeds the watermark where human speech carries the most energy, typically lower frequencies, due to auditory masking. 
This contrasts with WavMark, which places the watermark in higher frequency bands.
Embedding the watermark in lower frequencies is advantageous. 
For example, speech remains audible with a lowpass filter at 1500 Hz, but not with a highpass filter at the same frequency. 
This difference is measurable with PESQ in relation to the original audio, making it more beneficial to be robust against a lowpass filter at a 1500 Hz cut-off than a highpass filter at the same cut-off:

\begin{tabular}{cccc}
    Filter Type & PESQ & \ours & \wavmark \\
    \midrule
    Highpass 1500Hz & 1.85 \xmark & 0.7 & 1.0 \\
    Lowpass 1500Hz & 2.93 \cmark & 1.0 & 0.7 \\
\end{tabular}

\section{Experimental details}

\subsection{Loudness}\label{app:loudness}

Our loudness function is based on a simplification of the implementation in the torchaudio~\cite{yang2021torchaudio} library.
It is computed through a multi-step process. 
Initially, the audio signal undergoes K-weighting, which is a filtering process that emphasizes certain frequencies to mimic the human ear's response. 
This is achieved by applying a treble filter and a highpass filter.
Following this, the energy of the audio signal is calculated for each block of the signal. 
This is done by squaring the signal and averaging over each block. 
The energy is then weighted according to the number of channels in the audio signal, with different weights applied to different channels to account for their varying contributions to perceived loudness.
Finally, the loudness is computed by taking the logarithm of the weighted sum of energies and adding a constant offset. 

\subsection{Robustness Augmentations}\label{app:augmentations}

Here are the details of the audio editing augmentations used at train time (T), and evaluation time (E):
\begin{itemize}[itemsep=0pt, topsep=2pt, leftmargin=8pt]
    \item \textbf{Bandpass Filter:} Combines highpass and lowpass filtering to allow a specific frequency band to pass through.
    (T) fixed between 300Hz and 8000Hz; (E) fixed between 500Hz and 5000Hz.
    \item \textbf{Highpass Filter:} Uses a highpass filter on the input audio to cut frequencies below a certain threshold.
    (T) fixed at 500Hz; (E) fixed at 1500Hz.
    \item \textbf{Lowpass Filter:} Applies a lowpass filter to the input audio, cutting frequencies above a cutoff frequency.
    (T) fixed at 5000Hz; (E) fixed at 500Hz.
    \item \textbf{Speed:} Changes the speed of the audio by a factor close to 1. 
    (T) random between 0.9 and 1.1; (E) fixed at 1.25.
    \item \textbf{Resample:} Upsamples to intermediate sample rate and then downsamples the audio back to its original rate without changing its shape.
    (T) and (E) 32kHz.
    \item \textbf{Boost Audio:} Amplifies the audio by multiplying by a factor.
    (T) factor fixed at 1.2; (E) fixed at 10.
    \item \textbf{Duck Audio:} Reduces the volume of the audio by a multiplying factor.
    (T) factor fixed at 0.8; (E) fixed at 0.1.
    \item \textbf{Echo:} Applies an echo effect to the audio, adding a delay and less loud copy of the original.
    (T) random delay between 0.1 and 0.5 seconds, random volume between 0.1 and 0.5; (E) fixed delay of 0.5 seconds, fixed volume of 0.5.
    \item \textbf{Pink Noise:} Adds pink noise for a background noise effect.
    (T) standard deviation fixed at 0.01; (E) fixed at 0.1.
    \item \textbf{White Noise:} Adds gaussian noise to the waveform.
    (T) standard deviation fixed at 0.001; (E) fixed at 0.05.
    \item \textbf{Smooth:} Smooths the audio signal using a moving average filter with a variable window size.
    (T) window size random between 2 and 10; (E) fixed at 40.
    \item \textbf{AAC:} Encodes the audio in AAC format.
    (T) bitrate of 128kbps; (E) bitrate of 64kbps.
    \item \textbf{MP3:} Encodes the audio in MP3 format.
    (T) bitrate of 128kbps; (E) bitrate of 32kbps.
    \item \textbf{EnCodec:} Resamples at 24kHz, encodes the audio with EnCodec with $nq=16$ (16 streams of tokens), and resamples it back to 16kHz.
\end{itemize}
Implementation is done with the \texttt{julius} python library.

\subsection{Networks architectures (Fig.~\ref{fig:archs}) \label{app:arch}}

The watermark generator is composed of an encoder and a decoder, both incorporating elements from EnCodec~\cite{defossez2022high}. The encoder applies a 1D convolution with 32 channels and a kernel size of 7, followed by four convolutional blocks. Each of these blocks includes a residual unit and down-sampling layer, which uses convolution with stride $S$ and kernel size $K = 2S$. 
The residual unit has two kernel-3 convolutions with a skip-connection, doubling channels during down-sampling. The encoder concludes with a two-layer LSTM and a final 1D convolution with a kernel size of 7 and 128 channels. 
Strides $S$ values are (2, 4, 5, 8) and the nonlinear activation in residual units is the Exponential Linear Unit (ELU). 
The decoder mirrors the encoder but uses transposed convolutions instead, with strides in reverse order.

The detector comprises an encoder, a transposed convolution and a linear layer. 
The encoder shares the generator's architecture (but with different weights). 
The transposed convolution has $h$ output channels and upsamples the activation map to the original audio resolution (resulting in an activation map of shape $(t, h)$). 
The linear layer reduces the $h$ dimensions to two, followed by a softmax function that gives sample-wise probability scores.

\subsection{MUSHRA protocole detail}\label{app:MUSHRA}
The MUSHRA protocol is a crowdsourced test in which participants rate the quality of various samples on a scale of 0 to 100. 
The ground truth is provided for reference. 
We utilized 100 speech samples, each lasting 10 seconds. 
Each sample was evaluated by at least 20 participants.
As part of the study, we included a low anchor, which is a very lossy compression at 1.5kbps, encoded using EnCodec. 
Participants who failed to assign the lowest score to the low anchor for at least 80\% of their assignments were excluded from the study.
For comparison, the ground truth samples received an average score of 80.49, while the low anchor's average score was 53.21.

\subsection{Attacks on the watermark}\label{app:attacks}

\paragraph{Adversarial attack against the detector.}
Given a sample $x$ and a detector $D$, we want to find $x' \sim x$ such that $D(x') = 1 - D(x)$.
To that end, we use a gradient-based attack.
It starts by initializing a distortion $\delta_{adv}$ with random gaussian noise.
The algorithm iteratively updates the distortion for a number of steps $n$. 
For each step, the distortion is added to the original audio via $ x = x + \alpha . \mathrm{tanh} (\delta_{adv})$, passed through the model to get predictions. 
A cross-entropy loss is computed with label either 0 (for removal) or 1 (for forging), and back-propagated through the detector to update the distortion, using the Adam optimizer.
At the end of the process, the adversarial audio is $x + + \alpha . \mathrm{tanh} (\delta_{adv})$.
In our attack, we use a scaling factor $\alpha=10^{-3}$, a number of steps $n=100$, and a learning rate of $10^{-1}$. 
The $\mathrm{tanh}$ function is used to ensure that the distortion remains small, and gives an upper bound on the SNR of the adversarial audio.

\vspace{-0.2cm}
\paragraph{Training of the malicious detector.}
Here, we are interested in training a classifier that can distinguish between watermarked and non-watermarked samples, when access to many samples of both types is available.
To train the classifier, we use a dataset made of more than 80k samples of 8 seconds speech from Voicebox~\cite{le2023voicebox} watermarked using our proposed method and a similar amount of genuine (un-watermarked) speech samples. 
The classifier shares the same architecture as \ours's detector. 
The classifier is trained for 200k updates with batches of 64 one-second samples. 
It achieves perfect classification of the samples. 
This is coherent with the findings of Voicebox~\cite{le2023voicebox}.

\end{document}